\documentclass[fleqn,usenatbib]{mnras}
\usepackage{newtxtext,newtxmath}

\usepackage[T1]{fontenc}
\usepackage{lineno}

\DeclareRobustCommand{\VAN}[3]{#2}
\let\VANthebibliography\thebibliography
\def\thebibliography{\DeclareRobustCommand{\VAN}[3]{##3}\VANthebibliography}

\usepackage{graphicx}	
\usepackage{amsmath}	
\usepackage{xspace}

\def\psrb{PSR~B1259$-$63\xspace}
\def\psrbb{PSR~B1259$-$63/LS~2883\xspace}
\def\lat{\textit{Fermi}-LAT\xspace}
\def\hess{H.E.S.S.\xspace}

\begin{document}
\title[Multi-GeV PSR B1259-63 detection]{Multi-GeV Fermi-LAT Detection of PSR B1259-63}
\author[D. Malyshev et al.]{\parbox{\textwidth}
{
          D. Malyshev$^{1}$\thanks{Email:denys.malyshev@astro.uni-tuebingen.de}, 
          M. Chernyakova$^{2,3}$,
          A. Finn Gallagher$^2$,
          A. Kuzin$^1$,
          N. Matchett$^4$,
          A. Santangelo$^1$,\\ 
          Iu. Shebalkova$^2$,
          B. van Soelen$^4$\thanks{E-mail:vansoelenb@ufs.ac.za}
}
\\ \\ 
$^{1}$ Institut f{\"u}r Astronomie und Astrophysik T{\"u}bingen, Universit{\"a}t T{\"u}bingen, Sand 1, D-72076 T{\"u}bingen, Germany \\
$^{2}$ School of Physical Sciences and Centre for Astrophysics \& Relativity, Dublin City University, Glasnevin, D09 W6Y4, Ireland.\\
$^{3}$Dublin Institute for Advanced Studies, 31 Fitzwilliam Place, Dublin 2; \\
$^{4}$ Department of Physics, University of the Free State, PO Box 339, Bloemfontein 9300, South Africa \\
\\
}

\date{Received $<$date$>$  ; in original form  $<$date$>$ }

\label{firstpage}
\pagerange{\pageref{firstpage}--\pageref{lastpage}} 
\pubyear{2025}

\maketitle

\begin{abstract}
\psrbb is a classical gamma-ray binary detected from radio to TeV energies near periastron. Using over 17 years of Fermi-LAT observations, we report the first significant detection ($8\sigma$ in likelihood analysis) of the system in the 10-100 GeV energy range, over orbital phases from -400 to +100 days relative to periastron. The observed spectrum is well described by a power law with photon index $\Gamma = 1.9 \pm 0.1$ and shows a flux level consistent with that measured at TeV energies by H.E.S.S. The smooth connection between the Fermi-LAT and TeV spectra suggests that the detected multi-GeV emission traces the rising tail of the inverse-Compton component extending into the TeV regime. The presence of detectable emission hundreds of days before periastron indicates high-energy activity over a larger orbital phase range than previously established, enabling new constraints on particle-acceleration and radiative processes in the system.
\end{abstract}

\begin{keywords}
gamma rays: stars – pulsars: individual: PSR B1259-63 – binaries: general
\end{keywords}

\section{Introduction}

\psrbb is a well-studied gamma-ray binary system comprising of a rapidly rotating radio pulsar and a massive O9.5Ve Be star, LS~2883 \citep{Johnston1992, Negueruela2011}. The pulsar has a spin period of approximately 47.8 ms and orbits its companion in a highly eccentric orbit ($e \approx 0.87$) with an orbital period of $P_{\rm orb} = 1236.724526(6)$ days \citep{Shannon2014}. The epoch of periastron passage, $T_0 = 53071.2447290(7)$ MJD, is well established from timing measurements \citep{Shannon2014}, enabling precise phase-resolved studies of the system’s emission properties.

The multiwavelength emission of \psrbb arises from the interaction of the relativistic pulsar wind with the radiation field and the stellar wind (composed of a polar wind and a circumstellar decretion disk) of the Be star  \citep{1994ApJ...433L..37T, kirk99}. 

In the radio band, pulsed emission from the pulsar is eclipsed for several weeks around periastron due to free-free absorption and scattering in the dense circumstellar environment \citep{johnston96,johnston05}. Concurrently, unpulsed synchrotron emission arises from the shocked relativistic particles of the pulsar wind and shows a complex, orbitally modulated light curve in radio and X-ray bands~\citep{Connors2002, Chernyakova2024,Chernyakova2025}. 

X-ray observations with \textit{Suzaku}, \textit{XMM-Newton}, \textit{Swift}, \textit{INTEGRAL} and \textit{NuSTAR} reveal a characteristic, orbit-to-orbit reappearing, double-peaked light curve bracketing periastron, with two broad maxima occurring roughly $\sim 15$~days before and after periastron \citep{chernyakova06,Tam2011,Chernyakova2025, we_isgri}. A third distinct X-ray maximum was also detected after the 2021 periastron passage \citep{Chernyakova2021}. In the GeV band, \lat{} first firmly detected \psrbb up to a few GeV during the 2010 periastron passage \citep{Abdo2011}.
The GeV light curve exhibits complex temporal behavior: relatively low flux near periastron is followed by intense flares typically occurring 30--80 days after periastron \citep[see][and references therein]{Chernyakova2025}. These flares consists of a number of subflares that can be as short as few minutes and at these timescales reach gamma-ray luminosities by a factor of $\sim 30$ exceeding the pulsar’s spin-down power~\citep{Johnson2018}. Notably, no obvious correlation between the GeV emission and optical, X-ray or very high energy gamma-ray variability has been established to date \citep{2013A&A...551A..94H, Chernyakova2020}, suggesting distinct emission regions or mechanisms.

Optical observations of the Be star and its decretion disk reveal variations in line profiles and continuum flux related to the pulsar’s orbit \citep{vanSoelen2016}. However, no clear correlation between optical variability and X-ray or GeV emission has been found, underscoring the complexity of the system and possible time-dependent changes in the disk structure.

\begin{figure*}
    \centering
    \includegraphics[width=0.45\linewidth]{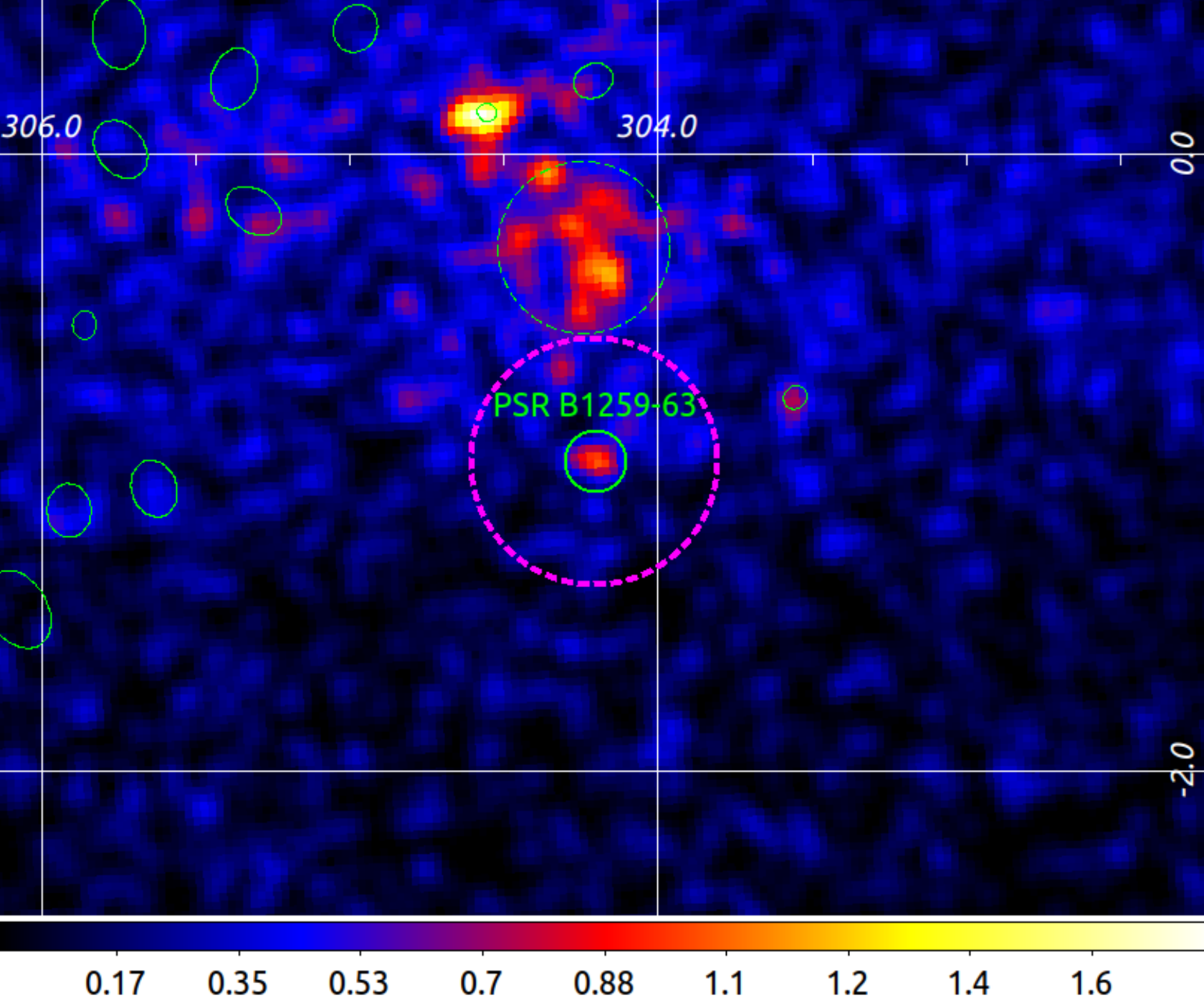}
    \includegraphics[width=0.51\linewidth]{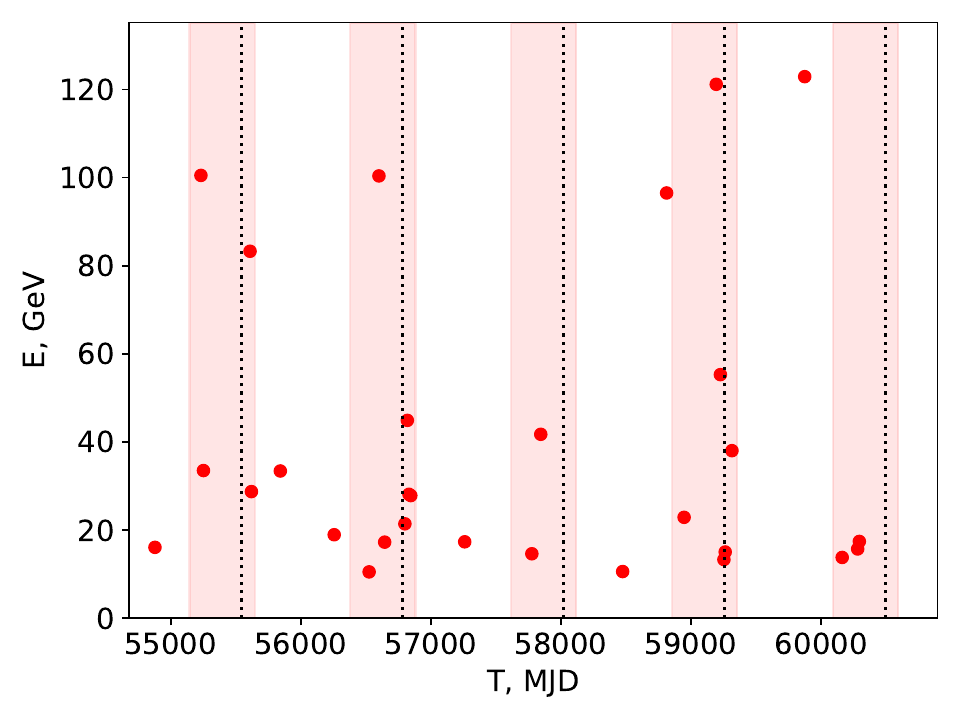}
    \caption{\textit{Left panel:} Count map of the \psrb region in galactic coordinates above 10~GeV smoothed with $0.1^\circ$ gaussian kernel. The green ellipses illustrate positions of nearby 4FGL catalogue sources. The thin green dashed circle stands for an extended source HESS~J1303-631. The solid green and dashed magenta circles correspond to the ON and OFF regions around the \psrb position. \textit{Right panel:} Arrival times of the photons above 10~GeV in $0.1^\circ$-radius circle around the \psrb position vs. the energy of the photons. Vertical dotted lines present the times of periastron, shaded regions illustrate $(-400; +100)$~days around the periastron.}
    \label{fig:arrival_times}
\end{figure*}

In the TeV regime, \hess{} observations have detected variable emission peaking near periastron with two broad maxima similar to the X-ray light curve \citep{Aharonian2005, Abdalla2018}.
Furthermore, \hess{} has detected TeV emission well before and well after periastron, from $-106$ to $+128$ days relative to $T_0$ \citep{Aharonian2005, Abdalla2018}. Whether this TeV emission detected over an extended orbital phase range reflects particle acceleration over a large spatial region or varying environmental conditions remains an open question.

Spectrally, the TeV emission is suggested to originate from the inverse-Compton (IC) radiation of the shocked relativistic electrons of the pulsar wind~\citep{Chernyakova2020} and is described by a power law with photon index $\Gamma \sim 2.7-2.8$ extending from several hundred GeV to multi-TeV energies \citep{Hess2024}. Extrapolation of this power law to lower energies significantly exceeds the observed GeV flux, implying the presence of a spectral break or low-energy cutoff in the TeV band near or below $\sim 100$\,GeV. Despite extensive observational efforts, no clear evidence for such a spectral break or even a variability of the spectral index has been found in the TeV band\footnote{The reported variability of the index $\Delta\Gamma\approx0.56\pm 0.18_{\rm stat}\pm 0.1_{\rm syst}$ corresponds to $\sim 2\sigma$ variability.}~\citep[see e.g.][]{Hess2024}. 

Interestingly, a strong correlation between X-ray and TeV fluxes was observed during the 2021 periastron passage, including the period of the third X-ray peak \citep{Hess2024}. This suggests a possible common origin of accelerated particles producing synchrotron X-rays and inverse Compton TeV photons during this phase. However, since such a correlation has not been observed in other periastron passages, it remains unclear if this relation is a persistent feature or unique to certain orbital cycles.

Despite decades of multiwavelength observations, the characterization of the rising part of the IC emission below $\lesssim 100$~GeV energies remained challenging due to limited photon statistics and instruments sensitivity. In this work, we present a detailed analysis of \lat{} data focusing on energies from 10\,GeV to more than 100~GeV, resulting in the first firm detection of \psrbb in this energy range from -400 to +100 days around the periastra. This study provides new constraints on the highest-energy emission processes in this archetypal gamma-ray binary and contributes to the understanding of particle acceleration and radiation in pulsar-Be star systems.

\section{Data Analysis}
\label{sec:data_analysis}

We analyzed more than 17 years of data collected by \lat between August 4, 2008, and August 6, 2025. To search for high-energy emission from \psrbb, we restricted the analysis to energies above 10~GeV, where \lat provides an excellent point-spread function\footnote{See, e.g., \href{https://www.to.infn.it/~maldera/Fermi-LAT/test_LATperf/lat_Performance.htm}{\lat performance webpage}} of $\sim 0.1^\circ$ and the astrophysical background is relatively low. The analysis was performed using the latest \texttt{fermitools} (v2.4.0) with the P8R3 instrument response functions for the \texttt{SOURCE} event class.

\begin{figure*}
\centering
\includegraphics[width=0.48\linewidth]{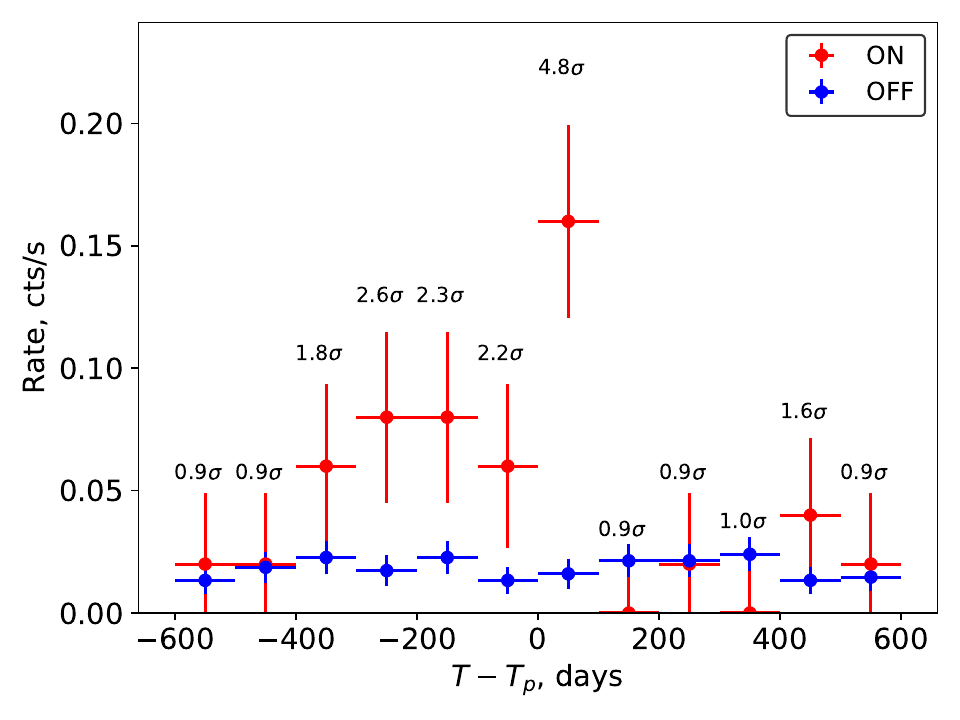}
\includegraphics[width=0.48\linewidth]{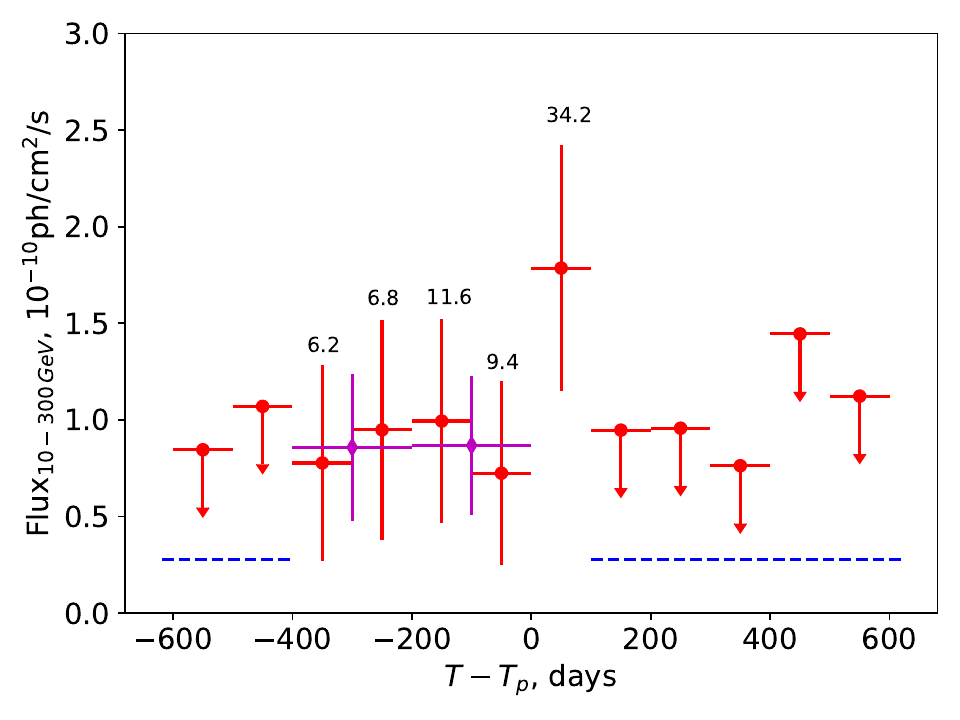}
\caption{Orbital light curve of \psrbb above 10GeV. \textit{Left:} Light curve from the ON–OFF analysis. Red points show the ON-source count rate within $0.1^\circ$ of the \psrb position. Blue points show the background (OFF) count rate in an annulus between $0.1^\circ$ and $0.4^\circ$. The significance above each point corresponds to the Poisson significance of $N_{ON}$ counts above the expected background. \textit{Right:} Orbital light curve from the likelihood analysis. The dashed blue line shows the combined flux upper limit around apastron (+100 to –400~days), while the magenta point indicates the average flux detected between –400 and +100~days. For points with $TS>1$, the corresponding TS values are shown above the markers.}
\label{fig:lc}
\end{figure*}

As a first step, we carried out a simple ON–OFF analysis~\citep[see, e.g.,][]{methods}. We filtered the \lat photons according to the standard procedure recommended by the \lat collaboration.\footnote{See \href{https://fermi.gsfc.nasa.gov/ssc/data/analysis/scitools/}{standard LAT filtering procedure}} Photons within a $0.1^\circ$ radius around \psrb (the ``ON'' region) were selected (see Fig.~\ref{fig:arrival_times}, left panel, for the count map). For the background (``OFF'' region), we selected photons in an annulus between $0.1^\circ$ and $0.4^\circ$. This yielded $N_{\rm ON}=29$ photons in the ON region and $N_{\rm OFF}=171$ photons in the OFF region.

Assuming the background photons are spatially uniform, the expected number of counts in the ON region is
\begin{equation}
M = N_{\rm OFF}\cdot\frac{A_{\rm ON}}{A_{\rm OFF}} = 11.4
\end{equation}
where $A_{\rm ON}$ and $A_{\rm OFF}$ are the areas of the ON and OFF regions, respectively. The Poisson probability of observing $N_{\rm ON}$ counts given an expectation of $M$ is
\begin{equation}
\label{eq:poisson}
P(N_{\rm ON} | M ) = \frac{M^{N_{\rm ON}} e^{-M}} {N_{\rm ON}!} \approx 5.7\times 10^{-6},
\end{equation}
corresponding to a $\sim 4.5\sigma$ detection of \psrbb above 10\,GeV in the 17-year time-averaged data. The arrival times and energies of ON photons are shown in Fig.~\ref{fig:arrival_times} (right panel, red points). Vertical dashed lines mark the periastron epochs, near which \lat emission up to a few GeV has previously been reported. The detected photons are not confined to a single short flare but span a broad time range and energies up to $\sim 123$~GeV. We interpret this as the maximum energy at which \psrbb is detected with \lat.

We next constructed an orbital-folded light curve using 13 bins of 100~days each, covering the interval from –600 to +600~days around periastron $T_p$. The left panel of Fig.~\ref{fig:lc} shows the count rate for ON (red) and OFF (blue) regions. Detection significances, computed using Eq.~\ref{eq:poisson}, are labeled above each point. For illustration, statistical uncertainties were estimated using $\Delta N = \sqrt{N+0.25}\, +\,1$ for low-count signals \citep[see, e.g.,][]{barlow04}. The figure shows that \psrbb is detected with $\sim 4.8\sigma$ significance in the $(0; +100)$~days bin and marginally ($\sim 2\sigma$) in four consecutive bins from –400 to 0 days before periastron. These detection intervals are indicated by red shaded regions in Fig.~\ref{fig:arrival_times}.

We note, that $4.8\sigma$ detection significance obtained in the ON-OFF analysis above can not serve as a firm estimation of the significance of the detection. Namely, the number of the source and background photons can be affected by  the presence of nearby diffuse source (see dashed green circle in Fig.~\ref{fig:arrival_times}); galactic-latitude intensity-gradients in the galactic diffuse emission; possible leakage of the photons from ON to OFF region due to the PSF of Fermi-LAT (only 68\% PSF containment at 10 GeV is $\sim 0.1^\circ$). Thus the significance value derived from the ON-OFF analysis  should be treated rather as an estimation and motivation for the more accurate 3D likelihood analysis.

\begin{figure*}
\centering
\includegraphics[width=0.49\linewidth]{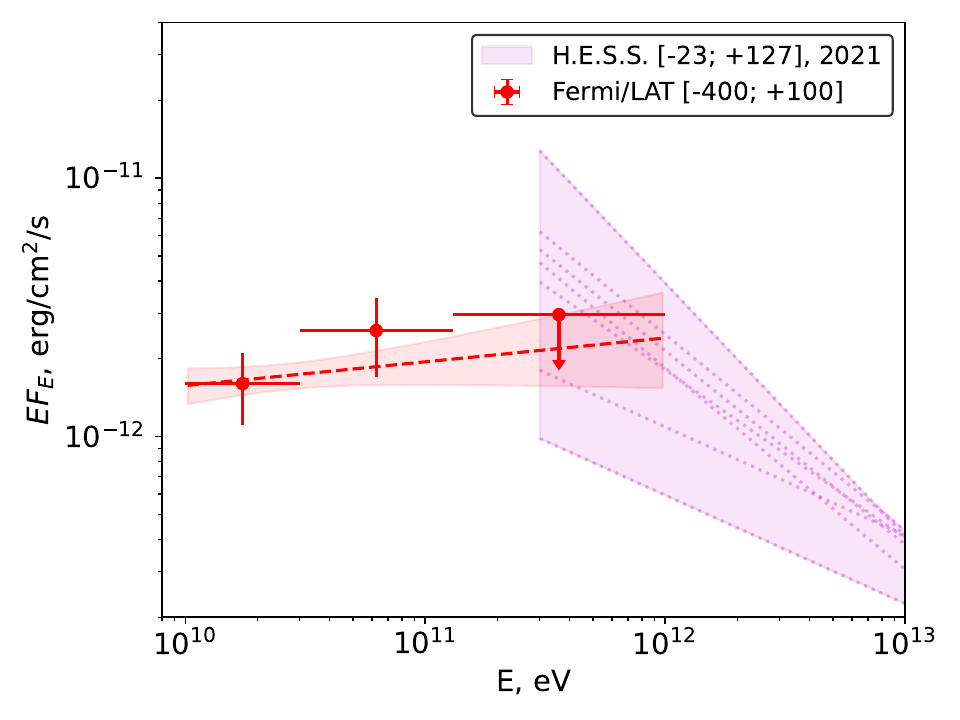}
\includegraphics[width=0.49\linewidth]{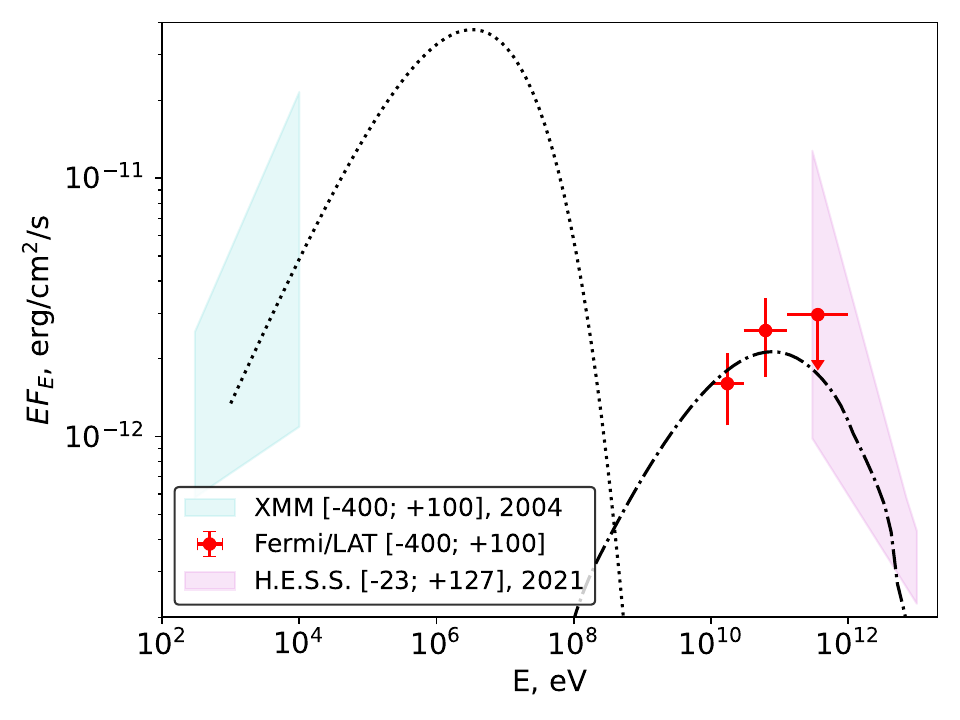}
\caption{Multiwavelength spectrum of \psrbb. \textit{Left:} Joint \lat spectrum (–400 to +100 days around periastron, averaged over 2008–2025) and \hess spectrum (–23 to +127 days around the 2021 periastron~\citet{Hess2024}). The red dashed line and shaded region show the best-fit power-law model to \lat data above 10~GeV and its $1\sigma$ statistical band. Dotted magenta lines show \hess spectra from different intervals in 2021. \newline \textit{Right:} Broadband spectrum from X-rays to TeV. The cyan shaded region indicates the flux range observed during –400 to +100 days around the 2004 periastron~\citep{chernyakova06}. Black dotted and dash-dotted lines show synchrotron and inverse-Compton components of a simple one-zone model (see text). }
\label{fig:spectrum}
\end{figure*}

We performed a 3D binned likelihood analysis of \lat data~\citep[see, e.g.,][]{methods}, fitting a spatial–spectral model of the $5^\circ$ region of interest (ROI). The model included all 4FGL DR4 \lat sources \citep{4fgl} and standard templates for Galactic and extragalactic diffuse backgrounds. To determine the flux of \psrbb, we fixed all source spectral parameters to their catalog values except for the normalizations, which were left free. Similarly to the ON–OFF approach, we analyzed the data in 100~day long bins around periastron (Fig.~\ref{fig:lc}, right). Upper limits (95\% C.L.) were derived using the \texttt{IntegralUpperLimit} module provided by \texttt{fermitools}. Test statistic (TS) values are labeled for bins with $TS>1$, with significance approximated by $\sigma\sim\sqrt{TS}$~\citep{mattox96}. 
The dashed blue lines shows the combined upper limit from periods without detection, i.e. from $+100$ to $+800$ days (equivalent to $+100$; $-400$~days).
Magenta points represent broader 200-day bins from –400 to 0 days, where \psrbb was detected with $TS=12$ (–400 to –200 days) and $TS=21$ (–200 to 0 days), corresponding to $3-4\sigma$ detections. Over the full –400 to +100 day interval, the likelihood analysis yields $TS\approx 64$, corresponding to $\sim 8\sigma$ detection.

For this period, (-400; +100)~days, we determined the spectral properties of \psrbb above 10~GeV. The spectrum between 10~GeV and 1~TeV is well described by a power law with index $\Gamma = 1.9\pm 0.1$ and flux $F_{10-130\ \mathrm{GeV}} = (1.0\pm 0.1)\times 10^{-10}$ph\,cm$^{-2}$\,s$^{-1}$ (Fig.~\ref{fig:spectrum}, left). The best-fit power law is shown as a dashed red line with the $1\sigma$ statistical butterfly from the covariance matrix. Dotted magenta lines show \hess spectra around the 2021 periastron for comparison.

A fit limited to 10–130~GeV yields a marginally harder slope $\Gamma=1.4\pm 0.3$, but the \lat statistics do not allow a firm conclusion on spectral curvature. A log-parabola fit provides only a weak improvement over the power law, with $-2\Delta\mathcal{L}\approx 2.8$, corresponding to $<2\sigma$ statistical preference for the spectral curvature.

\section{Results and Discussion}
\label{sec:discussion}

In the above, we presented an analysis of $>10$~GeV \lat data, resulting in the detection of \psrbb up to energies of $\gtrsim 100$~GeV. In the orbital phase–folded light curve (see Fig.~\ref{fig:lc}, right panel), the system is detected from $-400$ to $+100$~days around periastron with a significance of $\sim 8\sigma$. Furthermore, the source is detected well before periastron with a significance of $3$–$4\sigma$ in the intervals $(-400; -200)$~days and $(-200; 0)$~days. Between $(0; +100)$~days, \psrbb is detected with $TS \sim 34$, corresponding to a statistical significance of $5$–$6\sigma$.

No significant emission is detected in our analysis for the period from $+100$ to $+800$~days (equivalent to $-400$~days) after periastron. The detection significance increases toward periastron between $-400$ and $+100$~days, suggesting variable flux during this period. However, the limited sensitivity of \lat prevents firm conclusions.

Over the entire detection period ($-400$ to $+100$~days), the spectrum of \psrbb is well described by a power-law model with index $\Gamma = 1.9 \pm 0.1$ (see Fig.~\ref{fig:spectrum}, left panel). This spectrum is consistent with the flux level observed by \hess around the 2021 periastron passage, though it exhibits a harder spectral index. This suggests the presence of a spectral break or turnover near $\sim 100$~GeV energies. Note that, \lat data alone show only a marginal ($< 2\sigma$) preference for a curved spectrum (e.g., a log-parabola) over a simple power law.

A spectral break or turnover below $\sim 1$~TeV has been predicted by many models of the system’s emission \citep{Chernyakova2015, chen19, Chernyakova2025} and is supported by TeV observations. In particular, a simple power-law extension of the TeV spectrum into the GeV band is inconsistent with previously reported \lat upper limits~\citep{Hess2024}.

To illustrate how the observed spectral turnover can be explained, we consider an order of magnitude estimation for \psrbb where the multi-GeV and TeV emission results from inverse-Compton scattering of relativistic electrons, and the X-ray emission from their synchrotron radiation of the same population of electrons (Fig.~\ref{fig:spectrum}, right panel). In this figure, the \lat spectrum from the current work (averaged over 17~years, ($-400; +100$)~days around periastron) is shown as data points. The magenta shaded region shows the range of \hess spectra from $-23$ to $+127$~days around the 2021 periastron~\citep{Hess2024}, and the cyan region represents XMM-Newton spectra from $-400$ to $+100$~days around the 2004 periastron~\citep{chernyakova06}. Note that the TeV and keV data are not simultaneous with the multi-GeV data.

For the model, we adopt parameters which are similar to those that have previously been discussed in literature.
The electrons follow a cutoff power-law distribution with index $\Gamma_e = 1.8$ and cutoff energy $E_{\mathrm{cut},e} = 10$~TeV. The magnetic field in the emission region located at an average distance $d = 2$~a.u. from the Be star is $B_0 = 1$~G. The synchrotron and IC components are shown in Fig.~\ref{fig:spectrum} (right panel) as black dotted and dot–dashed curves, respectively. These components reproduce the data reasonably well, in line with earlier models. We therefore argue that the \lat detection above 10~GeV represents the low-energy part of the IC component, peaking at $\sim 100$~GeV and extending into the TeV band.

In this illustrative estimate, similar orbital variability patterns are expected in the 10–100~GeV and TeV energies. This is supported by \hess observations, which detected the system between roughly $-100$ and $+100$~days around periastron. Note, that no \hess observations have been reported for earlier epochs (prior to $-100$~days).

The detection of \psrbb at $\sim -400$~days before periastron raises important questions about the origin of the very-high-energy (VHE) emission. In the model of~\citet{Chernyakova2020}, the VHE emission is produced near the apex of the shock cone formed by the interaction of the stellar outflow with the relativistic pulsar wind. The apex’s position is set by the momentum balance between the two outflows, giving $d_{\mathrm{apex}} = \alpha \cdot D$, where $D$ is the separation between the Be star and the pulsar. For the orbital parameters from~\citet{Shannon2014}, $D$ varies from $\sim 10$~a.u. at $-400$~days, to $\sim 1$~a.u. at periastron, and $\sim 5$~a.u. at $+100$~days.

For spherically symmetric winds, $\alpha$ is independent of $D$ and depends just on the momenta of stellar and pulsar winds. In this model, the VHE emission is IC radiation from relativistic electrons scattering off stellar photons, with flux proportional to 
\begin{equation}
F\propto\alpha^{-2} D^{-2}t_{\rm cool,esc} \, \eta(D)
\end{equation}
where $\eta(D)$ is the electrons' acceleration efficiency (fraction of particles accelerated) and $t_{\rm cool,esc}$ is the minimum between electrons' radiative cooling times and escape (adiabatic) times~\citep[see e.g.][]{kirk99}. Close to periastron the cooling timescales for $\sim 1$~TeV electrons producing multi-GeV photons are as short as $\sim 10^3$~s \citep[for synchrotron losses, see e.g.][]{Chernyakova2020}. However, far from periastron these timescales can be substantially longer as the synchrotron cooling scales with binary separation as 
\begin{equation}
t_{\rm cool, sync}\propto B^{-2}\propto (1-\alpha)^{2}D^{2},
\end{equation}
assuming that the magnetic field strength drops linearly with the distance from the pulsar. Similarly, the IC cooling timescale will decrease with binary separation as,
\begin{equation}
t_{\rm cool, IC}\propto u_{\rm ph}\propto \alpha^2 D^2,
\end{equation}
where $u_{\rm ph}$ is the energy density of the target photons \citep[see e.g.][]{khangulian14}. The escape timescale
\begin{equation}
t_{\rm esc} = D/v_{\rm esc} \approx 5\cdot 10^3\,\mbox{s}\, (D/10\,\mbox{a.u.})(v_{\rm esc}/c)
\end{equation}
thus can be smaller or comparable to the radiative cooling time. Consequently, assuming constant acceleration efficiency $\eta(D)={\rm const}$, the observed flux scales as  
\begin{equation}
F\propto D^{-2}\min(t_{\rm cool}, t_{\rm esc})\propto D^{0\, .. -1}.
\end{equation}

Thus the expected ratio of GeV fluxes between $T_1 = -400$~days ($\sim 10$~a.u. binary separation) and $T_2 = 0$~days ($\sim 1$~a.u. separation) is $1-10$, roughly consistent with the observed factor of a few. We note, however, that the observed flux ratio is close to the middle of the estimated range, i.e.\ this suggests the transition from escape to cooling losses dominance at certain orbital phases and/or a non-constant acceleration efficiency $\eta(D)$ \citep[see also the discussion in][]{Hess2024}. Additionally, the assumption of a constant value of $\alpha$ may not be valid, as it could change if the pulsar-stellar wind interaction shifts from the Be star's polar wind to the equatorial disk.

The apparent time asymmetry of the emission around the periastron can be connected to the distortion or destruction of the stellar disc, leading to a more complicated shock structure and less efficient acceleration. Alternatively, the asymmetry can be explained by an anisotropic nature of the IC emission~\citep[see e.g.][]{kirk99}.

We therefore strongly encourage continued observations of \psrbb significantly before and after periastron with current facilities (e.g., \hess) and future instruments \citep[e.g. CTA;][]{cta}. 
Such data will enable detailed studies of flux and possible spectral variability at different orbital phases well before and after periastron, shedding light on the origin of the VHE emission from this system.

\section*{Acknowledgments}
The authors would like to thank the reviewer for their very helpful suggestions, which improved this paper. 
 The authors acknowledge support by the state of Baden-W\"urttemberg through~bwHPC. This work was supported by Deutsches Zentrum f\"ur Luft- und Raumfahrt e.V. (DLR) grant 50OR2409. BvS acknowledges support from the National Research Foundation of South Africa (grant number 119430). MCh and IuSh acknowledge support from the European Space Agency (ESA) in the framework of the PRODEX Programme (PEA 4000120711). The authors wish to acknowledge financial support from the Centre for Astrophysics and Relativity at DCU.

\section*{Data Availability}
The data underlying this article will be shared on reasonable request to the corresponding authors.

\def\aj{AJ}%
\def\actaa{Acta Astron.}%
\def\araa{ARA\&A}%
\def\apj{ApJ}%
\def\apjl{ApJ}%
\def\apjs{ApJS}%
\def\ao{Appl.~Opt.}%
\def\apss{Ap\&SS}%
\def\aap{A\&A}%
\def\aapr{A\&A~Rev.}%
\def\aaps{A\&AS}%
\def\azh{AZh}%
\def\baas{BAAS}%
\def\bac{Bull. astr. Inst. Czechosl.}%
\def\caa{Chinese Astron. Astrophys.}%
\def\cjaa{Chinese J. Astron. Astrophys.}%
\def\icarus{Icarus}%
\def\jcap{J. Cosmology Astropart. Phys.}%
\def\jrasc{JRASC}%
\def\mnras{MNRAS}%
\def\memras{MmRAS}%
\def\na{New A}%
\def\nar{New A Rev.}%
\def\pasa{PASA}%
\def\pra{Phys.~Rev.~A}%
\def\prb{Phys.~Rev.~B}%
\def\prc{Phys.~Rev.~C}%
\def\prd{Phys.~Rev.~D}%
\def\pre{Phys.~Rev.~E}%
\def\prl{Phys.~Rev.~Lett.}%
\def\pasp{PASP}%
\def\pasj{PASJ}%
\def\qjras{QJRAS}%
\def\rmxaa{Rev. Mexicana Astron. Astrofis.}%
\def\skytel{S\&T}%
\def\solphys{Sol.~Phys.}%
\def\sovast{Soviet~Ast.}%
\def\ssr{Space~Sci.~Rev.}%
\def\zap{ZAp}%
\def\nat{Nature}%
\def\iaucirc{IAU~Circ.}%
\def\aplett{Astrophys.~Lett.}%
\def\apspr{Astrophys.~Space~Phys.~Res.}%
\def\bain{Bull.~Astron.~Inst.~Netherlands}%
\def\fcp{Fund.~Cosmic~Phys.}%
\def\gca{Geochim.~Cosmochim.~Acta}%
\def\grl{Geophys.~Res.~Lett.}%
\def\jcp{J.~Chem.~Phys.}%
\def\jgr{J.~Geophys.~Res.}%
\def\jqsrt{J.~Quant.~Spec.~Radiat.~Transf.}%
\def\memsai{Mem.~Soc.~Astron.~Italiana}%
\def\nphysa{Nucl.~Phys.~A}%
\def\physrep{Phys.~Rep.}%
\def\physscr{Phys.~Scr}%
\def\planss{Planet.~Space~Sci.}%
\def\procspie{Proc.~SPIE}%
\let\astap=\aap
\let\apjlett=\apjl
\let\apjsupp=\apjs
\let\applopt=\ao
\bibliographystyle{aa}
\bibliography{biblio}
\end{document}